# Leveraging Geospatial Information to address Space Epidemiology through Multi-omics—Report of an Interdisciplinary Workshop

Annette L. Sobel[1]*, Kenneth Yeh[2], Elaine Bradford[2], Colin Price[2], Joseph Russell[2], Gene Olinger[2], Sheila Grant[1], Chi-Ren Shyu[1]

**AFFILIATIONS**
[1]*University of Missouri, Department of Electrical and Computer Engineering, Columbia, MO, USA*
[2]*MRIGlobal, Kansas City, MO, USA*
*Author to whom correspondence should be addressed:
   Annette Sobel
   bigbitbucket@mac.com

**ABSTRACT**

This article will summarize the workshop proceedings of a workshop conducted at the University of Missouri that addressed the use of multi-omics fused with geospatial information to assess and improve the precision and environmental analysis of indicators of crew space health. The workshop addressed the state of the art of multi-omics research and practice and the potential future use of multi-omics platforms in extreme environments. The workshop also focused on potential new strategies for data collection, analysis, and fusion with crosstalk with the field of environmental health, biosecurity, and radiation safety, addressing gaps and shortfalls and potential new approaches to enhancing astronaut health safety and security. Ultimately, the panel proceedings resulted in a synthesis of new research and translational opportunities to improve space and terrestrial epidemiology. In the future, early disease prevention that employs new and expanded data sources enhanced by the analytic precision of geospatial information and artificial intelligence algorithms.

## I. INTRODUCTION

With extended duration, potentially fast transit interplanetary space missions imminent, the primary objective of the space medicine community is keeping astronauts healthy, with particular emphasis on identification of pre-existing astronaut health conditions anticipating likely radiation hazards (especially radiation- induced cancers, acute radiation sickness, and trans-generational DNA and health effects), determining countermeasure strategies and susceptibility throughout space transit. [1] [2] [3] Other high probability human effects due to environmental exposure include central nervous system effects, cataracts, and infectious diseases. Although acute radiation exposure of 1-4Gy is symptomatic, the latency period for expression necessitates more immediacy of detection of cosmogenic ally generated neutrons and secondary effects. These methods include metabolomics and improved methods of detection of low level neutron fluxes such as neutron scintillator spectroscopy. [4] Currently, methods to estimate risk of human exposure and susceptibility are limited due to data availability.

By taking a systems biology approach, we may begin to understand the evolution of space pathogens and how they evolve genetically. The development of an open access biomedical information framework is required and involves collection of comprehensive astronaut data and organization into a massive taxonomy correlated to mission and environmental exposures. [5] https://genelab.nasa.gov/ecosystem. (This work is supported by the Translational Research Institute for Space Health through NASA Cooperative Agreement NNX16AO69A.)

Co-incident with space biomedical database development and data collection is precision medical instrumentation development for sampling and analysis. This paper will begin to address the role of multi- omics in the future of precision temporal and spatial data collection and space epidemiology, particularly from the perspective of precision health care in space and on earth.

(https://www.bcm.edu/academic-centers/space-medicine/translational-research-institute)

## II. MATERIALS AND METHODS

This thematic workshop was organized in the format of one keynote, two comprehensive panels, and a synthesis session. The primary objectives of the workshop were to (1) discuss the state of the art in global multi-omics platforms and current/future applications and potential funding sources of this technology platform in extreme environments The keynote speaker Mr. Ryan Toma from Viome®, https://www.viome.com provided an overview of the future of global omics platforms in a complex threat environment: strategies for data collection, analysis, and fusion. The first panel 'State of the art in global multi-omics and extreme environment' was a comprehensive recapitulation and expansion on the keynote.

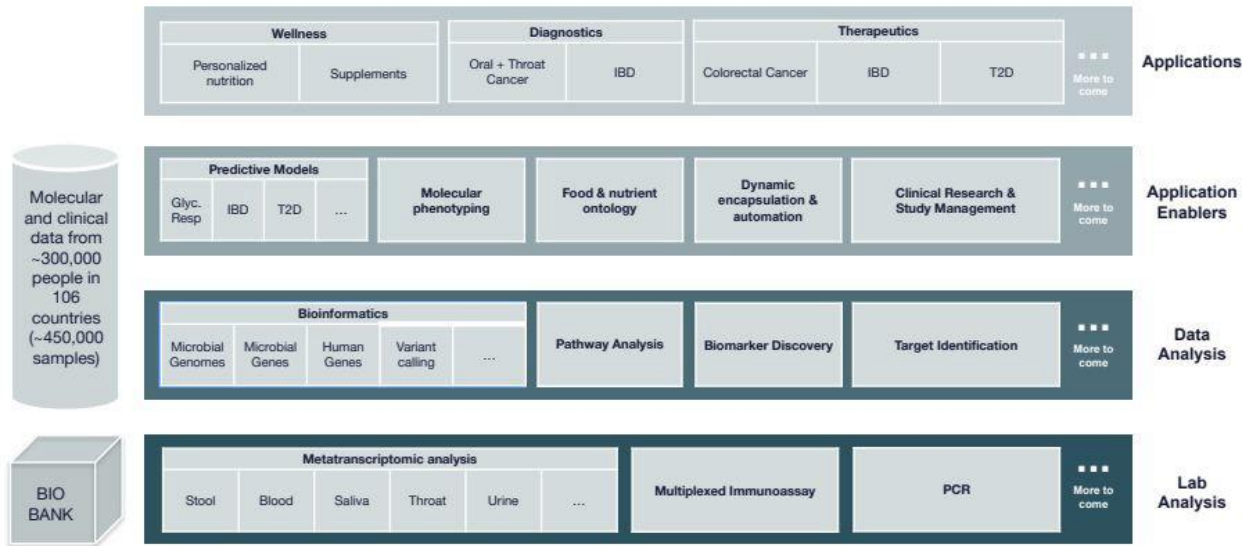

**Figure 1:** The ViOS® system architecture. Clinically validated metatranscriptomics.

Other systems were referenced and include Metabolon® https://www.metabolon.com/why-metabolomics/your-guide-chapter-1-metabolomics-metabolites-metabolome/. The second panel 'Geospatial information and biosecurity crosstalk: gaps and shortfalls' was intended to provided interdisciplinary perspectives on a primarily applied data-mining, analytic, and collection-oriented Field. "The future of global 'omics' platforms in the complex threat environment (of space): strategies for data collection, analysis, and fusion "describes the interdisciplinary panel that ensued.

**PANEL I. State of the art of multi-omics and extreme environments: changing the future of epidemiology for earth, space and beyond.**

Panelists: Joe Russell (MRI Global, focus area is applied biology and bioinformatics; Lyndon Coghill (MU, focus area is genetic markers of disease), Jimmy Wu (Translational Research Institute of Space Health, biomedical engineer), Ryan Toma (Viome ® research scientist)

Microbiome-human/animal interactions, chronic disease assessment and the impact of, personalized nutrition captures the general topic of the first session and supplementary panel discussion. The panel expanded upon the topic of the evolving future of epidemiology, the development of space models of biomarker evolution, and the process of returning biomarkers to normal/baseline levels. For example, radiation effects are observable acutely at a cellular level, but not symptomatically, and early intervention may prevent the ensuing cancer formation. The EXPAND database is designed to ingest a variety of biomedical, environmental, and mission data types and formats and will be fully manipulable for determination of environmental impacts and potential interventions to secure crew health in space. Of great interest is early identification and mitigation of crew radiation exposure profiles during interplanetary travel, coupled with development of bio- markers of individual radiation exposure and susceptibility of target organs/tissues. We understand that acute and chronic exposure to ionizing radiation IR) generates differing profiles based on induction of complex biological responses. An interesting review of current status and future directions of radiation metabolomics was published by Menon et. al. in Frontiers of Oncology [5].

A new paradigm is evolving (as expounded in Panel 2) that begins with targeting the biological architecture and includes development of high-fidelity models of components that will impact the individual and ambient environment. Subsequently, test and evaluation of modifications to environment and measurable impact will be performed to establish risk mitigation strategies. A major challenge will be improving signal: noise and time required for a significant data collection. For example, COVID trace-back to a Huanan Seafood Wholesale Market epicenter in Wuhan (https://www.science.org/doi/10.1126/science.abp8715) was substantiated as was the initial famous case pinpointing the source of the 1854 Broad Street Cholera outbreak in Soho, London attributed to Dr. John Snow (https://en.wikipedia.org/wiki/1854_Broad_Street_cholera_outbreak).

Collection methodologies have rapidly advanced to include the Vios® system and to take biological samples, perform RNA analysis, utilize AI/ML tools for categorizing and taxonomy and potentially, 'tree-of-life' mapping, biomarker discovery, target identification, and development of targeted diagnostics and therapeutics for measurably improved patient outcomes. Although a small cohort (current n=600), the following general areas are being explored: nutritional therapies/ precision supplements, oral/throat cancer, infectious diseases, colorectal cancer, autoimmune disease vaccines, MCI/Alzheimer's and chronic dementia, pancreatic cancer, and targeted drug discovery/development.

An initial step in effective treatment requires an understanding of the chemistry of disease to identify disease transition before clinical indicators are present. Some cancers/noninfectious diseases are examples of present areas of study.

RNA sequencing gives functional results, measurably better than DNA, which can only do 'risk' of disease. In addition, RNA is modulated with diet and exercise, making the data more operationally relevant.

**PANEL II. Geospatial information and biosecurity crosstalk: Gaps and Shortfalls**

Panelists: Ram Raghavan (MU-focus area is epidemiology); Pal Palaniappan (MU-focus areas are information processing, computer engineering, aerial drone imaging, microscopy);

Grant Scott (MU-focus areas are computer engineering and geospatial analytics); Gene Olinger (MRI Global-focus areas are leading versus lagging indicators, Intelligence Community considerations).

There are limited datasets in spaceflight (and multi-omics), which complicates the utility of ML, especially deep learning applications. One approach is to anticipate and model spaceflight scenarios and how they might modify the interactive parameters of living systems with the environment. An example is the movement of pathogens in a microgravity, low humidity, low ultraviolet environment or during ambient radiation exposure and solar wind conditions (https://en.wikipedia.org/wiki/Solar_wind)

**Biocontainment and how it related to space/ extreme environments.**

Through his expertise working in BL3 and 4's, Gene Olinger, the focus is on reducing exposures as much as possible. MRIGlobal is moving into VR/ AR to supplement training [of high-risk incidents in low-risk training environments]. Relatedly, when you bring space samples back you don't want to contaminate them with Earth markers. As previously experienced, there are extensive extraterrestrial contamination potentials, dictating the importance of sample integrity, system of systems analysis and probability of contamination and exposure.

By zooming out, both literally on a geospatial scale and conceptually, and using tools such as AI, ML, and aerial data collection technologies, we will be able to make epidemiological connections and disease predictions previously unseen. Generally, the more data the better – and the more specific that data is to the question being asked, the better; especially as trends and connections arise that are not visible or understood until much later. Collecting data with future capabilities and insight in mind will set us up to answer questions yet undefined. Creating reliable, open databases from which researchers can reliably pull information is critical; this would reduce collection redundancies and aid in concurrent data specificity to the research questions being asked. There are many applications to this concept across personalized medicine, epidemiology/ biosecurity, disease evolution, and improving human health and disease prevention in space.

## III. RESULTS

**Revolutionizing nutrition**

Extreme environments demand a data-driven approach to nutritional assessment that is characterized by the following: highly personalized, and focused on disease prevention, mitigation, and health maintenance. Fad dietary supplements and approaches ignore the nuance of the individual. One approach is to develop a personalized model through human-assisted AI/ML training, and models that classify blood sugar spikes as anomalies. Subtle changes and associated metadata may be detected through clinically-experience human assistance. In synchrony, the biomarkers in the microbiome may be identified and then the targeted foods and supplements may be mediated. Although currently exploratory, this approach may be used in Crohn's/IBS/Depression/Anxiety, unbiased/unsupervised learning, aging targets, oral/throat cancers, and others. Identification of target samples and understanding individual susceptibility may be challenging and currently there is an 'art' to sample collection heavily dependent upon expertise.

**Shaping the future of epidemiology in extreme environments**

Anticipating complex medical care during space missions, NASA has developed a Medical Extensible Dynamic Probabilistic Risk Assessment Tool (MEDPRAT) [6] to help quantify medical components of spaceflight health risk due to hazard exposure such as radiation. Integration of space epidemiology data, especially for extended duration space missions, will be anticipatory of the unknown scientific questions stemming from challenging mission environments with a wide span of probability of exposure and individual vulnerabilities and the response of complex systems biology.

Artificial intelligence/machine learning algorithms (AI/ML) are ideal for analysis large, dynamic data sets of complex systems, and when coupled to multi-omics, create a powerful toolkit for analysis, 'sense-making' and derivation of population/individual susceptibility from aggregate and personal health data/metadata. However, when metadata such as geospatial/temporal information is used, caution must be applied to insure data standardization and precision. Additionally, scaling and extrapolations leading to population conclusions which must address confidentiality, ethical, cultural, and political concerns.

Geospatial information and multi-omics may help shape the future of space and exploration of extreme and threat (manmade and natural) environments. For example, understanding the dynamics of (re-)emerging infectious diseases especially in confined spaces and extreme environmental conditions depends on understanding the source and temporal evolutionary history of the genome, i.e., its 'tree of life'. This approach will naturally assist with pathogen countermeasures, reverse- threat engineering, and forecasting. Subsequently, a deeper understanding of specific biological triggers or biomarkers, and how they are modulated by the environment, is an emerging field coined 'meta-crypt-omics' by the research community of interest. This approach will lead to new therapeutic options to therapeutic- diagnostics.

An equally valuable and insightful tool for epidemiology is data and information visualization, particularly when used to reveal what is missing and to develop a holistic understanding of the existing/emerging ecosystem. Geospatial-temporal information is ideally suited for this approach to information display of actionable, intuitive intelligence/information. These reconfigurable databases will have greater utility with AI/ML tools.

**Recommendations**

With the advent of readily available commercial space travel, the opportunity and need to ensure crew health maintenance and disease prevention is imperative. Multi-omics is an adaptable, specific indicator platform which will enable situation awareness of comprehensive sets of environmental effects due to radiation, toxins, pathogens, atmospheric contaminants that may be unanticipated, potentially preventable, and reversible.

Additionally, this capability may help identify threat countermeasures at early stages of specific bioeffects. This workshop has attempted to provide a 'state of the art' assessment and review of promising translational research applications and tools which will likely benefit many future generations of space travelers.

Currently, major impediments to data analysis in space includes small collection samples and lack of significant heterogeneity of data samples. The hope is for near- to mid-term larger samples that are representative of human, animal, and plant populations to fully address systems biology implications of extreme environments and unanticipated stress-response effects that may alter our understanding and management of radiation exposure in cancer therapeutics, especially during pre-syndromic periods. In addition, the American Society for Radiation Oncology (ASTRO) Board of Directors supports the study of tumor genomics and epigenetics in advancement of improved precision medicine using radiation therapy. [7]

Rapidly advancing international translational research (in space and terrestrial settings) in multi- omics coupled to large scale data analytic and visualization tools will lead to readily manipulable data sets and mission-specific applications. Additionally, when analyzed in the context of geospatial information, space epidemiology and forecasting algorithms will become more reliable and precise indicators of extreme environment effects.

**IV. CONFLICT OF INTEREST STATEMENT**

The authors declare that the research was conducted in the absence of any commercial or financial relationships that could be construed as a potential conflict of interest.

**V. AUTHOR CONTRIBUTION STATEMENT**

All authors contributed strongly to content and organization. AS was the primary topical organizer and instigator as well as space medicine expert. SG,CRS facilitated the meeting and organizational approval, reviewed the manuscript and SG added reference and innovative insights. KY significantly assisted with content, organization, administrative approvals and SME speaker recruitment. JR, GO,CP, EB contributed to technical content, organization, and visionary applications/way forward. MV and RT are technical experts in metabolomics and platform technology development and created ViOS (R) architecture figure and explanatory comments.

## VI. CONTRIBUTION TO THE FIELD

This workshop identified a number of areas for expanded current and future research investment to ensure astronaut and space traveler health. Specifically, we identified the field of multi-omics as one offering unbounded potential in early identification of health vulnerabilities and predictive analysis of environmental effects. In addition, findings are translatable to personalized health care on earth. The spectrum of environmental challenges in space is virtually limitless and will offer profound opportunities for research in metabolic, psychological and zoonotic health, not to mention disease prevention. We also identified new opportunities for parallel work in expanded applications of AI and decision-assisting tools for health care and engineering professionals.

## VII. ETHICS STATEMENTS

*Studies involving animal subjects:* No animal studies are presented in this manuscript.

*Studies involving human subjects:* No human studies are presented in this manuscript.

*Inclusion of identifiable human data:* No potentially identifiable human images or data is presented in this study.

*Data availability statement:* Publicly available datasets were analyzed in this study. This data can be found here: no data link is available; this is a methodology report and synthesis of an emerging field by interdisciplinary SMEs.